\documentclass[prb,cha,twocolumn,superscriptaddress,preprintnumbers,showpacs, amsmath,amssymb]{revtex4-1}
\usepackage{bm}
\usepackage[colorlinks=true,linkcolor=blue,citecolor=blue]{hyperref}
\usepackage{amsmath}
\usepackage{amssymb}
\usepackage{amsthm}
\usepackage{amsfonts}
\usepackage{enumerate}
\usepackage{latexsym}
\usepackage{ifpdf}
\usepackage{graphicx}
\usepackage{makeidx}
\expandafter\ifx\csname package@font\endcsname\relax\else
 \expandafter\expandafter
 \expandafter\usepackage
 \expandafter\expandafter
 \expandafter{\csname package@font\endcsname}
 \fi
\usepackage{color}
\usepackage{times}

\begin{document}

\title{Anomalous  orbital  structure in two-dimensional titanium dichalcogenides}

\author{Banabir Pal}
\email {bp435@physics.rutgers.edu}
\affiliation{Department of Physics and Astronomy, Rutgers University, Piscataway, New Jersey 08854, USA}

\author{Yanwei Cao}
\email {ywcao@nimte.ac.cn}
\affiliation{Department of Physics and Astronomy, Rutgers University, Piscataway, New Jersey 08854, USA}
\affiliation{Ningbo Institute of Materials Technology and Engineering, Chinese Academy of Sciences, Ningbo, Zhejiang 315201, China}

\author{Xiaoran Liu}
\affiliation{Department of Physics and Astronomy, Rutgers University, Piscataway, New Jersey 08854, USA}

\author{Fangdi Wen}
\affiliation{Department of Physics and Astronomy, Rutgers University, Piscataway, New Jersey 08854, USA}

\author{M. Kareev}
\affiliation{Department of Physics and Astronomy, Rutgers University, Piscataway, New Jersey 08854, USA}

\author{A. T. N'Diaye}
\affiliation{Advanced Light Source, Lawrence Berkley National Laboratory, Berkeley, California 94720, USA}

\author{P. Shafer}
\affiliation{Advanced Light Source, Lawrence Berkley National Laboratory, Berkeley, California 94720, USA}

\author{E. Arenholz}
\affiliation{Advanced Light Source, Lawrence Berkley National Laboratory, Berkeley, California 94720, USA}

\author{J. Chakhalian}
\affiliation{Department of Physics and Astronomy, Rutgers University, Piscataway, New Jersey 08854, USA}

\date{\today}

\begin{abstract}
Generally, lattice distortions play a key role in determining the ground states of materials. Although it is well known that trigonal distortions are generic to most two dimensional transition metal dichalcogenides, the impact of this structural distortion on the electronic structure has not been understood conclusively. Here, by using a combination of polarization dependent X-ray absorption spectroscopy (XAS), X-ray photoelectron spectroscopy (XPS) and atomic multiplet cluster calculations, we have investigated the electronic structure of titanium dichalcogenides TiX$_2$ (X=S, Se, Te), where the magnitude of the trigonal distortion increase monotonically from S to Se and Te. Our results reveal the presence of an anomalous and large crystal filed splitting. This unusual kind of crystal field splitting is likely responsible for the  unconventional electronic structure of TiX$_2$ compounds. Our results also indicate the drawback of the distorted crystal field picture in explaining the observed electronic ground state of these materials and emphasize the key importance of metal-ligand hybridization and electronic correlation in defining the electronic structures near Fermi energy.
\end{abstract} 

\maketitle
The realization of numerous exotic electronic phases of graphene \cite{RMPgraphene, ref2graphene, ref3graphene} and the relentless  tendency to miniaturization of silicon-based electronics\cite{ref4silicon} have ignited  exhaustive research in a wide range of two dimensional (2D) layered materials. As a result, 2D transition metal dichalcogenides have emerged as the promising platform with intriguing electronic ground states and high potential for applicability in the field of microelectronics\cite{ref4Amicro,ref4Bmicro}, nanophotonics\cite{ref5nanophoto, ref6nanophoto2}, optoelectronics\cite{ref7opto, ref8opto2} and photovoltaics\cite{ref8Aphoto, ref8Bphoto, ref8Cphoto} to  name a few. A generic feature of this dichalcogenide family is the presence of structural distortions which play a critical role in defining the electronic ground state of these systems. Specifically, lattice deformations in chalcogenides invariably alter the interatomic interaction strength and thereby produces various novel electronic phases  including charge density wave in VSe$_2$\cite{ref8DVSe2,ref8EVSe2}, NbSe$_2$\cite{ref8FNbSe2} and TaSe$_2$\cite{ref8GTaSe2,ref8HTaSe2}, superconductivity  in FeSe$_{1-x}$Te$_x$\cite{ref8J}, insulating ground states in ReSe$_2$\cite{ref8JJ}, Weyl semi-metallic phase in MoTe$_2$\cite{ref8JJJ} and  more. Naturally, a proper understanding of the lattice distortions and their impact on the electronic structure are essential for  deterministic control of the rich physical properties of these 2D systems.\\ 
\indent Recently, titanium dichalcogenides TiX$_2$ (here X = S, Se and Te) have attracted significant attention of the  community as potential candidates for energy storage applications\cite{ref9intro,ref9Aintro,ref9Bintro,ref10intro2} due to the easily achievable lithium intercalation. While the usual crystallographic form of TiX$_2$ is the layered CdI$_2$ type\cite{ref11crystal}, these systems also possess a distinct trigonal distortion from an ideal octahedral crystal environment\cite{ref11crystal, ref12crystal2}. Previously, detailed structural investigation established that the magnitude of the distortion varies monotonically from nearly octahedral in  TiS$_2$\cite{ref11crystal} to highly distorted in TiTe$_2$\cite{ref12crystal2}. In addition, large body of work on transition metal compounds suggest that these generic trigonal distortions may have strong influence in modifying the energy level and electronic structure of the chalcogenides. The problem is vividly  illustrated by the case of TiSe$_2$ which undergoes the transition into a chiral charge density wave (CDW)\ state\cite{ref12CDW27, ref12CDW28} and further into a conventional CDW resulting in dramatic renormalization of electronic and structural properties. The fundamental challenge here is to separate the many-body effects associated with  excitonic condensate\cite{ref12CDW6, ref12CDW7, ref12CDW8} from the Jahn-Teller like instability from the strong-electron phonon coupling\cite{ref12CDW9, ref12CDW10}. In addition, recent extensive theoretical  work based on LDA+U\cite{ref12CDW11} along with the detailed ARPES \cite{ref12CDW34} study hinted on the importance of lattice distortions coupled with strong electron-electron correlations to explain the ground state properties of TiSe$_2$. Based on those findings and given the obvious importance of these the whole TiX$_2$ family, there have been several experimental and theoretical investigations to understand the impact of trigonal distortion on the electronic structure in these systems\cite{refprev1, refprev2, refprev3}. 
\begin{figure*}[h!t]
\begin{center}
\includegraphics[width=2.0\columnwidth]{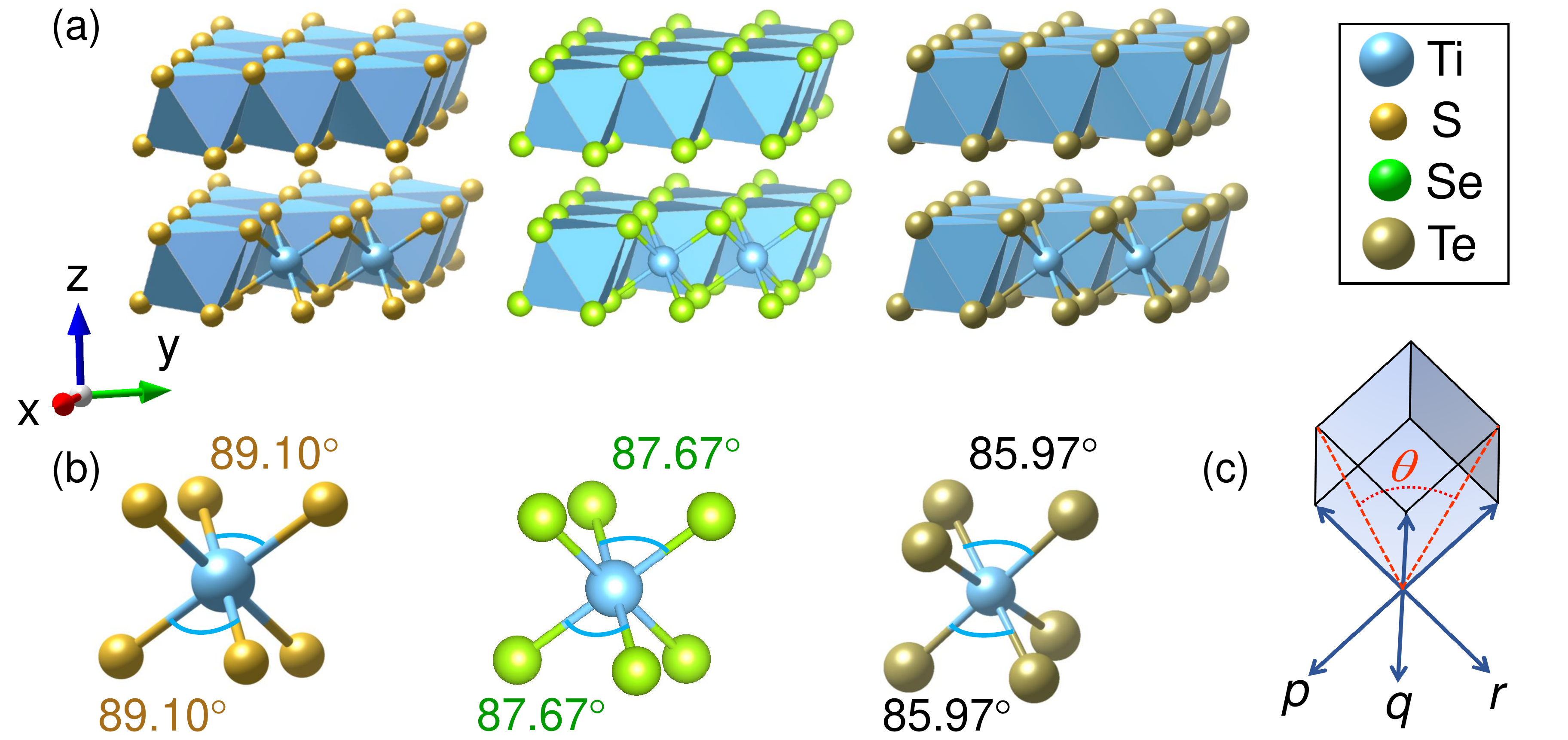}
\caption{(a) Schematic crystal structure of the TiS$_2$, TiSe$_2$, and TiTe$_2$ compounds showing nearly octahedral environment of each TiX$_2$ unit. (b) Distortion of the ideal octahedral environment across the chalcogenide series. It is to be noted that the distortion does not change the bond length but alters the bond angle. (c) Schematic representation of the trigonal distortions.}
\label{fig1}
\end{center}
\end{figure*}
However, in absence of systematic spectroscopic investigations corroborated with theoretical calculations the consequence of the trigonal distortion on the electronic structure of the TiX$_2$ systems has not convincingly been reported thus far. Here, we have addressed this issue by investigating the whole  family of TiX$_2$ single crystals by means of polarization dependent X-ray absorption spectroscopy in conjunction with the multiplet cluster calculations.  Our results unambiguously demonstrate the failure of the standard ionic configuration and distorted crystal field picture in predicting electronic ground states and thereby reveal the key importance of metal-ligand hybridization between titanium and chalcogen ions and electronic correlations  in defining their electronic properties.\\  
\indent As shown in Fig. 1(a), all members of the TiX$_2$ family crystallize into trigonal CdI$_2$ type layered structure with space group $P_{3m1}$ \cite{ref11crystal} . The crystal structure consists of repeated tri-layers (Fig. 1(a)) along $z$ direction; each tri-layer contains a titanium layer sandwiched between two layers of chalcogenides. Although the interactions between titanium and chalcogenides are strong within a tri-layer, the chalcogen bonding between two tri-layers is weak and  dominated by the van der Waals type interaction. Each tri-layer further experiences an elongated trigonal distortion where all six Ti-X bond length remains constant but X-Ti-X bond angle deviates from an ideal 90$^{\circ}$ in such a manner that the crystallographic lattice parameter along $z$ direction increases in length. As illustrated
in Fig. 1(b) the strength of elongated trigonal distortion increases monotonically from TiS$_2$ to TiTe$_2$  with three different X-Ti-X bond angles. The magnitude of such elongated trigonal distortion can be  estimated from $c/a$ ratio where $c$ and $a$ represent the corresponding unit cell parameter along $z$ and $x$ direction, respectively;   For an ideal octahedral environment, the $c/a$ ratio is close to 1.633 whereas the $c/a$ ratio for TiS$_2$, TiSe$_2$ and TiTe$_2$ are found to be 1.726, 1.732 and 1.808, respectively. This type of structural distortion can also be alternatively explained in terms of the distortion angle $\Theta$ between the diagonals of the $pq, qr, rp$ plane as schematically shown  in Fig 1(c). For a regular octahedron $\Theta=$  60$^{\circ}$ and  becomes lesser and greater than 60$^{\circ}$ for elongated and compressive trigonal distortions, respectively.\\  
\indent Chemical  quality, and the absence of chalcogen vacancies critical for\ most dichalcogenides are verified by X-ray photo electron spectroscopy (XPS) measurements  carried out on freshly cleaved TiX$_2$ single crystals to rule out the presence of any anion vacancies.  
\begin{figure*}[h!t]
\begin{center}
\includegraphics[width=1.95\columnwidth]{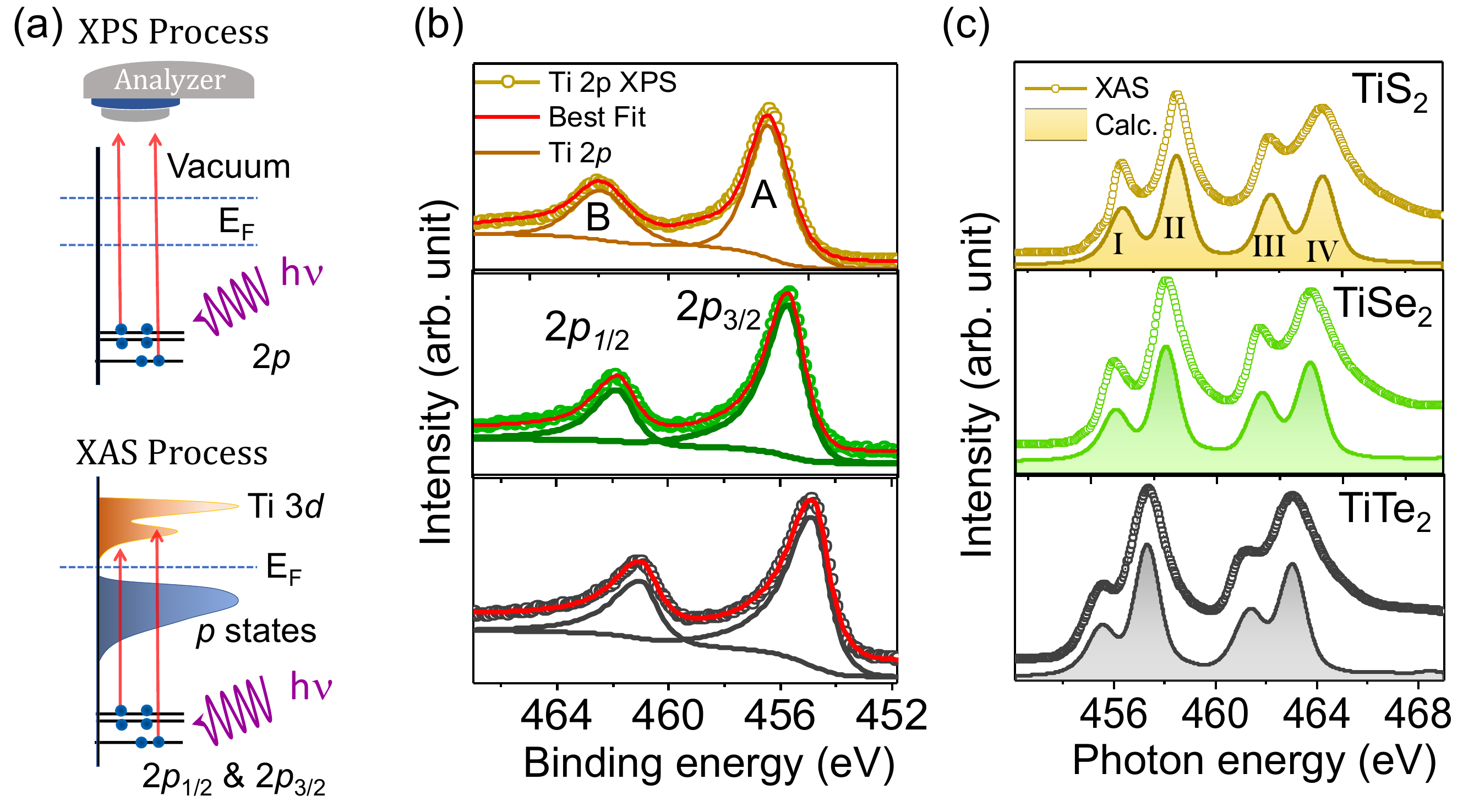}
\caption{(a) Schematic representations of the X-ray photoelectron (top) and X-ray absorption (bottom) spectroscopy process. (b) X-ray photoelectron spectra of Ti $2p$ core level. Each spectrum was decomposed with Lorentzian function convoluted with Gaussian functions. (c) Experimental XAS spectra (open circle) of three different TiX$_2$ systems were compared with atomic multiplet calculations (shaded area).}
\label{fig2}
\end{center}
\end{figure*}
The basic process associated with the XPS technique is shown schematically in Fig. 2(a). Generally, the chemical shift associated with each core level  provides important information about the charge state of the different elements of the system under investigation. Figure 2(b) displays typical Ti 2$p$ core level spectra for three different TiX$_2$ systems (Core level XPS spectra of Chalcogenides are shown in supporting information). Each Ti 2$p$ core-level spectrum shown in  Fig.2(b) is composed of two intense spin-orbit split doublet (marked as A and B) with a spin-orbit splitting strength close to 5.8 eV. Feature A appearing at a binding energy range between 454 eV to 459 eV represents Ti 2$p_{3/2}$ like states whereas feature B arises mainly from Ti 2$p_{1/2}$ like states. Interestingly, a systematic shift was found in moving from higher to lower binding energy in Ti 2$p$ core level spectra from TiS$_2$ to TiTe$_2$, respectively. These shifts in the binding energy were previously attributed to the reduction in ionic contribution in Ti-X chemical bond formations \cite{ref13xas}. In this work, each spectrum was decomposed using a Gaussian-Lorentz line profile and a Shirley type background function and can be accounted  within a single Gaussian-Lorentz line profile.The absence of the multiple peak structure in Ti 2$p$ core level spectra clearly implies the presence single valence Ti, and rules out any signature of anion vacancies in our TiX$_2$ samples.\\
\indent To investigate the evolutions in electronic structure across the TiX$_2$ series, XAS measurements were carried out on Ti L$_{2,3}$ edge at beamline 4.0.2 of the Advanced Light Source, at Lawrence Berkeley National Laboratory. In a typical Ti L$_{3,2}$ XAS process (see bottom of Fig. 2(a)), electrons are excited from Ti 2$p$ core level (from 2$p_{3/2}$ and 2$p_{1/2}$) to the Ti 3$d$ conduction band; each L$_{3,2}$ XAS spectrum splits into two edges ( L$_2$ and L$_3$ ) due to the spin-orbit coupling of Ti 2$p$ states. 
\begin{figure*}[h!t]
\begin{center}
\includegraphics[width=1.4\columnwidth]{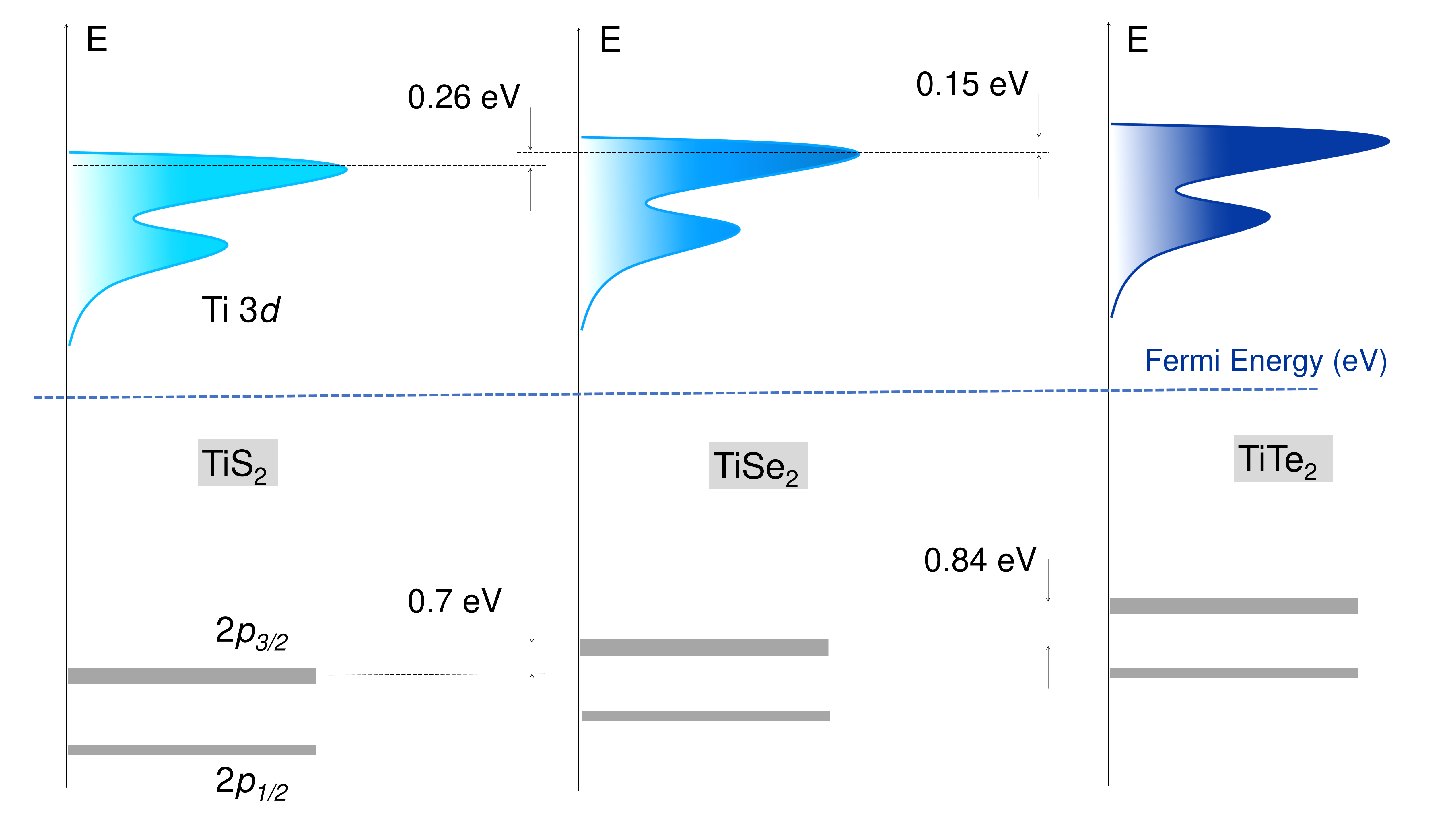}
\caption{(a) Band alignment of 2$p$ and 3$d$ states in TiS$_2$, TiSe$_2$ and TiTe$_2$ as obtained from X-ray photoelectron and X-ray absorption spectroscopy.}
\label{fig3}
\end{center}
\end{figure*}

Experimentally obtained Ti L$_{2,3}$ XAS spectra for three different TiX$_2$ systems are shown in Fig. 2(c) as open circles. As seen both L$_3$ and L$_2$ peaks of each spectrum exhibit a double hump feature arising primarily from nearly octahedral crystal field effects which splits the fivefold degenerate 3$d$ orbitals of Ti into doubly degenerate $e_g$ and triply degenerate $t_{2g}$ orbitals. For a quantitative understanding, theoretical atomic multiplet cluster calculations were performed within a TiX$_6$ cluster with O$_h$ point group symmetry and Ti$^{4+}$ ionic configuration. \cite{ref16} The obtained calculated spectra are shown as shaded line in Fig 2(c) confirm that only one types of charge state Ti$^{4+}$is  present  in our samples. In good agreement  with the  XPS spectra, we observed a gradual shift in the XAS spectra from TiS$_2$ to TiTe$_2$. This monotonic spectral shift suggests a systematic decrease in ($U_{dd}-U_{pd}$) from TiS$_2$ to TiTe$_2$ where $U_{dd}$ symbolizes the onsite coulomb interaction strength and $U_{pd}$ defines the core hole interaction potential. This  result indicates that in case of Te, the strongest interatomic $p-d$ interaction strength arises from the markedly increased metal-ligand hybridization. The summary of the relative energy position of the occupied 2$p$ states and unoccupied 3$d$ states of titanium obtained from XPS and XAS measurements is given schematically in Fig. 3. \\
\indent Despite high XAS\ sensitivity to charge state and local symmetry 
\begin{figure*}[b!t]
\begin{center}
\includegraphics[width=1.9\columnwidth]{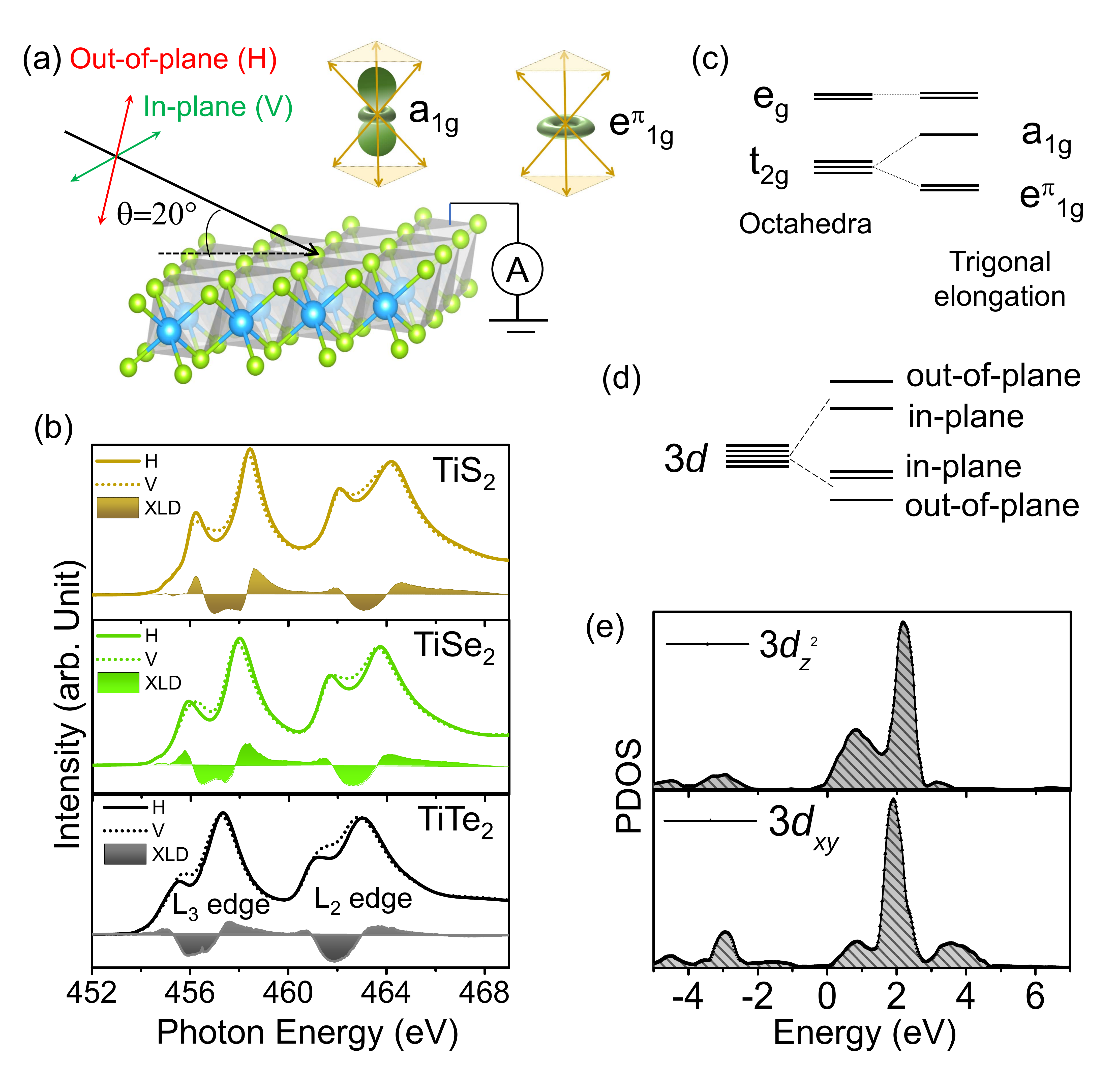}
\caption{(a) Experimental setup of the in-plane and out-of-plane polarization dependent XAS measurements. (b) Ti L$_{3,2}$ edge XAS/XLD spectra of three different TiX$_2$ compounds at room temperature.(c) Conventional theoretical model of an elongated trigonal distortion with $t_{2g}$ sub-band splitting. (d) Experimentally observed Ti 3$d$ sub-band splitting in TiX$_2$. (e) PDOS of 3$d_{z^2}$ and 3$d_{xy}$/3$d_{x^2-y^2}$ obtained from previous theoretical calculations\cite{ref18} on TiS$_2$ system where $x, y, z$ represent global co-ordinates axis as shown schematically in Fig. 1(a)} 
\label{fig4}
\end{center}
\end{figure*} 
it is still very challenging to capture the effect of trigonal distortions from non-polarized XAS measurement. To further explore the impact of trigonal distortion, detailed polarization dependent XAS measurements were carried out. The polarization dependent XAS process can probe local orbital character depending on their a relative orientation of orbitals with  respect to  crystallographic axes; for example it can be  sensitive to the sub-band splitting of $t_{2g}$ states which may emerge from local lattice distortions. The polarization dependent results  (in-plane vs. out-of-plane)  are schematically shown in Fig. 4(a). Since our samples are aligned along their natural [111] direction, it is expected that out-of-plane polarization will be sensitive to the orbitals  oriented along [111] direction whereas the in-plane polarization will probe those states which are oriented perpendicular to the [111] directions. Both out-of-plane (H) and in-plane (V) polarization dependent spectra are shown in Fig. 4(b). As immediately seen, a clear difference between the in- and out-of-plane XAS spectra is present which in turn implies the presence of distinct \textit{d}-orbital anisotropy. The corresponding XLD signal or the difference between the in- and out-of-plane XAS\ spectrum is shown as the shaded area in Fig. 4(b) for all three samples.\\
\indent To make a connection between the XLD results and orbital occupation we discuss electronic structure of the TiX$_2$ compounds. In the purely ionic limit, the electronic structure of these systems comprises of fully occupied $p$ states of chalcogen atom forming valence bands and unoccupied conduction bands with the predominant Ti 3$d$ character. Due to the effect of octahedral crystal field, five-fold degenerate 3$d$ orbitals of Ti split into a doubly degenerate higher energy $e_g$ and a triply degenerate lower energy $t_{2g}$ states\cite{ref14}. In addition, as shown schematically
in Fig. 4(c) due to elongated trigonal distortion\cite{ref15} $t_{2g}$ states further split into a higher energy singlet $a_{1g}$ and lower energy doublet $e^{\pi}_g$ states. The wave function of $a_{1g}$ and $e^{\pi}_g$ states is a linear combination of $d_{xy}$, $d_{yz}$ and $d_{xz}$ orbitals and can be  expressed as as $|a_{1g}\rangle$=$\frac{1}{\sqrt{3}}$$(|xy\rangle$+$|yz\rangle$+$|xz\rangle)$ and $|e^{\pi}_{g\pm}\rangle$=$\pm$$\frac{1}{\sqrt{3}}$ ($|xy\rangle$+$e^{\mp2i\pi/3}$$|yz\rangle$+$e^{\pm2i\pi/3}$$|xz\rangle$) \cite{ref17}. The description of the  mixed $a_{1g}$ and $e^{\pi}_g$ states becomes simplified  when one assumes \textit{z} axis along the [111] direction as shown in Fig. 1(a). The wave function of the $a_{1g}$ state has a similar shape along [111] direction as the $d_{z^2}$ orbital  along [001]; $e^{\pi}_g$ states are  perpendicular to the [111] directions and oriented in $xy$ plane (see the shape of the $a_{1g}$ and $e^{\pi}_g$ orbital Fig 4.(a))\\
\indent Based on this picture, one would naively expect that in-plane polarization will largely probe  $e^{\pi}_g$ states which are at lower energy  compared to the higher energy $a_{1g}$ state that are more sensitive to out-of-plane polarization. Contrary to  this expectation, the experimental spectra for all three systems are in complete variation with the conventional picture. More specifically, in the case of $t_{2g}$ states labeled by feature I and III in Fig 2(c), the out-of-plane polarization (shown by solid line in Fig. 4(b)) exhibits a lower energy peak position compared to the in-plane polarization (shown by dotted line). This implies the stabilization of $a_{1g}$ as the lowest occupied  orbital contrary to  the expected  $e^{\pi}_g$ state. This is completely opposite to what one would infer from the standard crystals field splitting arguments. The crystal field inversion picture is schematically shown in Fig. 4(d) (Quantitative Sub-band splitting of each TiX$_2$ compound has been shown schematically in supporting information). These findings clearly suggest that purely  ionic  picture which is normally very efficient in capturing the excitation spectra of Ti$^{4+}$ derived states in oxide systems is inadequate in explaining the origin of XLD signal in  dichalcogenides. Several factors might be responsible for such discrepancy in the description of the XAS spectra. Because of  the markedly  enlarged \textit{  p}-orbitals in S, Se and especially  Te,  unlike oxides, covalency may  play a dominant role in transition metal dichalcogenides. The synergetic effect of high metal-ligand hybridization and significant ligand field can  strongly affect the local electronic description of these systems.\\
\indent Adding more to the surprising finding, the crystal field splitting gap (10 $Dq$) between $t_{2g}$ and $e_g$ is found to be strongly dependent on the direction of polarization of the incident light. In particular, out-of-plane polarization (solid line in Fig. 4(b)) shows a higher crystal field gap compared to in-plane polarizations (dotted line in Fig. 4(b)). This  is unexpected since for the  trigonal distortion, the $e_g$ states should not experience any sub-band splitting. However, the energy position of the $e_g$ state is found to depend strongly on the types of polarization. All these discrepancies clearly establish that conventional crystal field picture fails to explain  the observed features in these system contrary to the  Ti$^{4+}$ based oxide materials.\\
\indent To resolve this inconsistency, the XAS spectra were compared with calculated projected partial density of states (PDOS) of 3$d_{z^2}$ and 3$d_{xy}$ states of TiS$_2$ \cite{ref13xas, ref18} as shown in Fig. 4(e). Interestingly, the calculated PDOS of 3$d_{z^2}$ of titanium exhibit a double hump structure analogous to the experiments when the systems were probed with out-of-plane polarization and  PDOS of the 3$d_{xy}$/3$d_{x^2-y^2}$ states (these two states are degenerate) also shows a double hump feature similar to in-plane polarization. Moreover, the energy gap between the doublet features is greater in the case of 3$d_{z^2}$ PDOS as compared to 3$d_{xy}$, and 3$d_{x^2-y^2}$ PDOS which is remarkably similar to the experimentally observed trend. The calculated PDOS is qualitatively consistent with experimental observations and  explains the appearance of the all features and their energy position as observed from experiments. The key feature of the theoretical calculation is the inclusion of strong metal-chalcogen hybridization which is responsible for the observed crystal field inversion and allows to capture the details of electronic structure  within the band-like description. This result implies that indeed covalency and metal-ligand hybridization are critical for the electronic properties of the TiX$_2$ family of materials.\\     
\indent In conclusion, detailed polarization dependent XAS measurements were carried out to understand the effect of trigonal lattice distortion on the electronic structure of the TiX$_2$ family. All systems were characterized  to rule out the presence of anion vacancies in these compounds. XLD spectra demonstrate the failure of conventional crystal field arguments in explaining the observed experimental features implying the crucial importance of covalency/metal-ligand hybridization in defining the electronic structure. Orbital projected DOS\ successfully reproduced the spectral features  observed in our experiments. The excellent  agreement between theory and  the XAS spectra  suggests the importance of band-like description in order to explain the electronic structure of transition metal dichalcogenides.      \\

\textbf{Experimental Section:}\\
\textbf{XPS measurements:} Ti 2$p$, S 2$p$, Se 3$d$ and Te 3$d$ core level XPS measurements were carried out in a lab-based Thermo Scientific X-ray Photoelectron Spectrometer furnished with a monochromatic Al $K_{\alpha}$ photon source and a hemi-spherical analyser with total energy resolution close to 0.45 eV. The base pressure of the main chamber was below 2$\times$10$^{-8}$ mbar during XPS measurements. All three TiX$_2$ compounds were mechanically exfoliated just before XPS measurements to avoid any surface contamination. The bulk sensitivity of the measurements was increased by collecting the ejected photoelectrons in a surface normal geometry. Photon energy of the source was calibrated using C 1$s$ core level spectra with a characteristic peak at around 284.6 eV binding energy. Each core level spectrum was decomposed in a casaXPS software using Gaussian-Lorentz type line profile.\\
\textbf{XAS /XLD measurements:} 
XAS/XLD measurements on three TiX$_2$ compounds were carried out at Ti L$_{3,2}$ edge at beamline 4.0.2 of the Advanced Light Source (ALS), at Lawrence Berkeley National Laboratory, USA. To avoid surface contamination, each single crystal was exfoliated mechanically in nitrogen atmosphere just before carrying out XAS measurements. No further surface treatments were performed as like vacuum heating or sputtering to clean the surface as these techniques could produce chalcogen vacancy in our systems. Total electron yield (TEY) detection technique were used during the XAS/XLD measurements.   \\

\textbf{Acknowledgement}: J.C., Y.C. and B.P. are supported by the Gordon and Betty Moore Foundation EPiQS Initiative through Grant No. GBMF4534. X.L. and F.W. are supported by the U.S. Department of Energy (DOE) under Grant No. DOE DE-SC 00012375 for synchrotron work and bulk crystal characterization. This research used resources of the Advanced Light Source, which is a Department of Energy Office of Science User Facility under Contract No. DE- AC0205CH11231. \\


\date{\today}

\end{document}